%% file: main.tex
\documentclass[10pt, conference, letterpaper]{IEEEtran}
\IEEEoverridecommandlockouts

\usepackage{color}                                       
\usepackage{xcolor}
\usepackage{graphicx}
\usepackage[labelformat=simple]{subcaption}
\usepackage{xspace}
\usepackage{multirow}
\usepackage[ruled,vlined]{algorithm2e}
\usepackage{ulem}
\usepackage{url}
\usepackage{cite}
\usepackage{float}
\usepackage{booktabs}
\usepackage{amsmath}
\usepackage{textcomp}
\usepackage{enumerate}
\usepackage{booktabs}
\usepackage{array}
\usepackage{tabularx}
\usepackage{pdfpages}
\usepackage{siunitx}
\usepackage{tabularx}
\usepackage{threeparttable}

\usepackage[colorlinks = false]{hyperref}
\hypersetup{
    draft,
    pdfborder={0 0 0}
}
\renewcommand{\abstractname}{\textbf{ABSTRACT}}
\begin{document}

\title{A Large-Scale IPv6-Based Measurement of the Starlink Network}

\author{
    Bingsen~Wang\IEEEauthorrefmark{1}\IEEEauthorrefmark{1},
    Xiaohui~Zhang\IEEEauthorrefmark{2}\IEEEauthorrefmark{1},
    Shuai~Wang\IEEEauthorrefmark{2}, 
    Li~Chen\IEEEauthorrefmark{3}, 
    Jinwei~Zhao\IEEEauthorrefmark{4}, 
    Dan~Li\IEEEauthorrefmark{2}, 
    Yong~Jiang\IEEEauthorrefmark{1}
    \\
    \thanks{\IEEEauthorrefmark{1}Bingsen Wang and Xiaohui Zhang contributed equally to this work and are co-first authors.}
    \IEEEauthorblockA{
        \IEEEauthorrefmark{1}Tsinghua Shenzhen International Graduate School
        \IEEEauthorrefmark{2}Tsinghua University\\
        \IEEEauthorrefmark{3}Beijing University of Posts and Telecommunications
        \IEEEauthorrefmark{4}University of Victoria\\
        wbs22@mails.tsinghua.edu.cn,
         zhangxh@mail.zgclab.edu.cn,
         \{wangshuai, lichen\}@zgclab.edu.cn,\\
         clarkzjw@uvic.ca,
         tolidan@tsinghua.edu.cn,
         jiangy@sz.tsinghua.edu.cn
    }
}








\maketitle

\begin{abstract}

Low Earth Orbit (LEO) satellite networks have attracted considerable attention for their ability to deliver global, low-latency broadband Internet services. In this paper, we present a large-scale measurement study of the Starlink network, the largest LEO satellite constellation to date. We first propose an efficient method for discovering active Starlink user routers, identifying approximately \num{5.98} million IPv6 addresses across \num{208} regions in \num{165} countries. Compared to general-purpose IPv6 target generation algorithms, our router-centric approach achieves near-complete coverage and, to the best of our knowledge, yields the most comprehensive known set of active IPv6 addresses for Starlink user routers. Based on the discovered user routers, we further propose an efficient method for mapping the Starlink backbone network and uncover a topology consisting of \num{49} Points of Presence (PoPs) interconnected by \num{98} links. We conduct a detailed statistical analysis of active Starlink user routers and PoPs, and further characterize the IPv6 address assignment strategy adopted by the Starlink network. Finally, we analyze the latency of Starlink user routers, propose a method to distinguish different types of users within the same region using \textit{outside-in} measurement, and identify the ongoing V2 Mini satellite deployment as a potential driver of the performance improvements. The dataset of the Starlink backbone network is publicly available at \url{https://ki3.org.cn/\#/starlink-network}. 


 
\end{abstract}


\input{01-INTRODUCTION}

\input{02-BACKGROUND}

\input{03-METHODOLOGY}

\input{04-EXPERIMENTS}

\input{05-DISCUSSION}

\input{06-CONCLUSION}

\section*{Acknowledgment}



We thank Prof. Jianping Pan (University of Victoria) and the anonymous reviewers for their valuable comments and constructive suggestions, which helped improve the clarity and presentation of this paper. This work was supported in part by the National Key R\&D Program of China (Grant No.~2022YFB3105000) and the National Natural Science Foundation of China (NSFC) (Grant No.~62502472). Shuai Wang and Jinwei Zhao are co-corresponding authors.




\bibliographystyle{IEEEtran}
\bibliography{refs}

\end{document}

%% file: 01-INTRODUCTION.tex
\section{INTRODUCTION}
Low Earth Orbit (LEO) satellite networks have shown tremendous potential in recent years as a supplement and extension to traditional terrestrial networks. LEO satellite networks are designed to provide high-speed, low-latency Internet services with global coverage. The wide coverage benefits communication in remote areas. Moreover, LEO satellite networks show a great advantage over terrestrial communication in latency, even in well-connected areas. The work in~\cite{handley2018delay} has shown that LEO satellite networks can provide lower latency communication than any possible terrestrial optical fiber network when communicating over a distance of \num{3000} km. 

Among existing deployments, Starlink, operated by SpaceX, stands out due to its rapidly growing constellation of more than \num{9000} satellites. As of December 2025, Starlink has connected over \num{9} million people with high-speed Internet in more than \num{155} countries, territories, and other markets~\cite{k1}. Despite its scale and global impact, Starlink’s internal architecture and operational mechanisms remain largely opaque, leaving the research community with only a limited understanding of how this large-scale LEO network is structured and operated.


Currently, some knowledge about Starlink comes from related simulations~\cite{kassing2020exploring,lai2020starperf,lai2023starrynet}, which are limited due to certain assumptions during the simulation process. Some studies have begun to investigate the network topology of Starlink based on real-world measurement in recent years~\cite{pan2023measuring, pan2024measuring, izhikevich2024democratizing}. The work in~\cite{izhikevich2024democratizing} probes publicly exposed Internet services in IPv4 from the outside to investigate the Starlink network latency, and measures over \num{2400} users across \num{27} countries. However, it covers only a small portion of Starlink's global users.  Similarly, on the censys.io platform~\cite{censys_as14593, censys_as45700}, only around \num{48000} active Starlink users' IP addresses are discovered, which is far fewer than the reported \num{9} million Starlink users by December 2025~\cite{k1}. The work in~\cite{pan2023measuring, pan2024measuring} investigates Starlink's access and backbone network on IPv4. Starlink’s IPv4 network extensively employs Carrier-Grade NAT (CGNAT), which means that most Starlink user routers lack public IPv4 addresses. This limits the capability to conduct large-scale \textit{outside-in} statistical analysis of Starlink networks from the Internet. On the other hand, the work in~\cite{pan2023measuring} also relies on the \textit{inside-out} approach to map the global backbone topology of Starlink, which requires at least one probe associated with each PoP to verify the results. Although RIPE Atlas\cite{ripe_atlas} is the largest network measurement platform, as of December 29, 2025, there are only around \num{100} active probes connected to the Starlink network, covering only 35 countries, making it challenging to map a complete global backbone topology.

\begin{figure*}[t] 
	\centering 
	\includegraphics[width=0.9\textwidth]{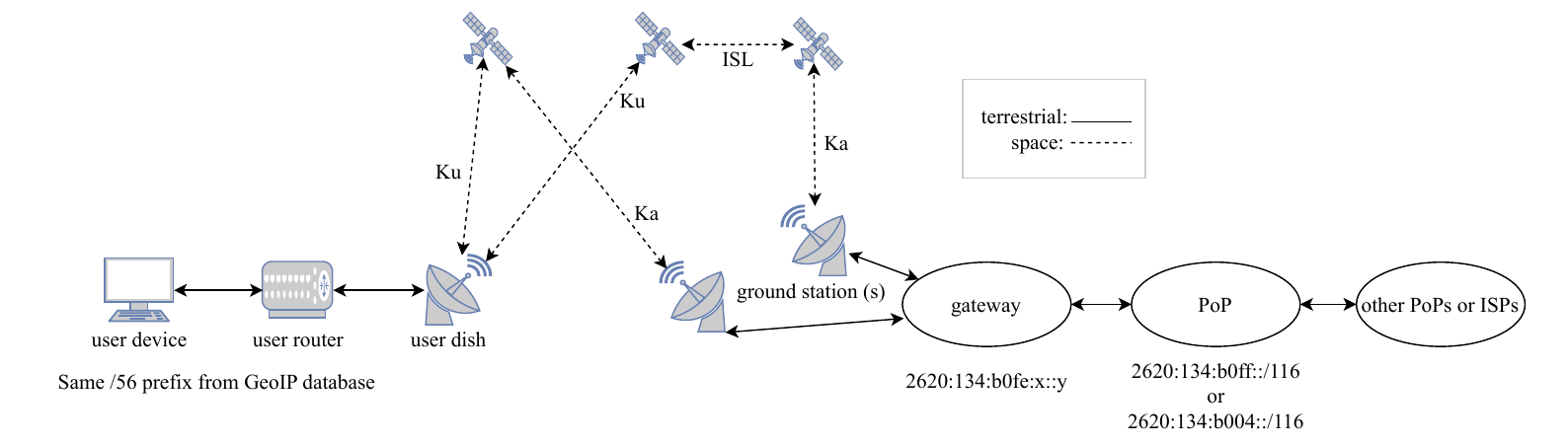} 
	\caption{Components of the Starlink network.} 
	\label{fig: Topology of the Starlink network} 
\end{figure*}



Fortunately, Starlink has already supported native IPv6 on all routers, although enabling it is not mandatory~\cite{starlink_ipv6_support}. Therefore, we propose leveraging public IPv6 addresses as targets for probing. Specifically, we utilize the IPv6 prefixes from the Starlink GeoIP Feed~\cite{p1} as input for our active IPv6 address scanning system to identify active Starlink user routers. Our method is capable of exhaustively enumerating all possible IPv6 addresses of Starlink user routers, thus providing more comprehensive coverage than any existing IPv6 target generation approaches. Unlike these general-purpose algorithms, which may yield addresses belonging to different devices of the same user and introduce unnecessary redundancy, our approach specifically targets user routers to achieve more precise and representative coverage. To the best of our knowledge, our method discovers the most comprehensive list of active IPv6 addresses of Starlink user routers, i.e., approximately \num{5.98} million addresses across \num{208} regions in \num{165} countries, as of December 29th, 2025. We use ``region'' to describe the geographic segmentation as defined in the Starlink GeoIP Feed. This usage remains consistent throughout the text except where explicitly noted.
This method not only provides probing targets for investigating the Starlink backbone network but also enables more accurate statistical analysis of Starlink users.

Considering that only data packets with both source and destination addresses within the Starlink network are forwarded between Starlink PoPs within the Starlink backbone network, we adopt the \textit{inside-out} approach with probes connected to Starlink network to investigate the Starlink backbone network. We use \num{6} Starlink dishes (including 2 probes from RIPE Atlas) to perform traceroute measurements to Starlink's public IPv6 addresses. We determine the associated PoP of an IP address using the mapping of IP prefixes to PoPs released by Starlink in~\cite{p2} and use the traceroute path to deduce the connections between Starlink PoPs. Using this method, we identify a total of \num{49} PoPs, which represent all PoPs currently in operation~\cite{peeringdb, googlemaps_starlink_pops}. These PoPs form the Starlink backbone network with \num{98} connections. Compared to the backbone network revealed by Pan et al.~\cite{pan2023measuring}, we achieve comparable coverage using an order of magnitude fewer Starlink dishes as probes, without requiring a probe to be located in every PoP.

Our statistical analysis shows that \num{49}\% of Starlink user routers are located in North America. Based on Starlink's prefix-to-PoP mapping, we quantify the number of routers served by each PoP. Interestingly, while the Starlink Miami PoP (mmmiflx1) serves only \num{225712} routers, significantly fewer than the Starlink Dallas PoP (dllstxx1)'s \num{321065} routers, it serves \num{19} regions, more than any other PoP in North America. This highlights the disparity in user distribution. Compared to dllstxx1, which has a more concentrated user base, mmmiflx1 covers a wider range of regions.

Furthermore, we utilize the massive dataset of active Starlink user routers to conduct latency measurements and uncover the mechanisms underlying the latency jitter of Starlink. Compared to prior work~\cite{izhikevich2024democratizing}, our study covers a broader range of geographic regions and a more diverse set of users. Through our analysis, we observe notable disparities in service quality among users within the same region, which could potentially be distinguished using our proposed variance-based classification method. Moreover, we observe a clear improvement in Starlink’s latency performance by comparing measurement results at different times and suggest that the ongoing V2 Mini satellite deployment is a potential driver for these gains.



The main contributions are summarized as follows:
{\sloppy 
\begin{itemize}
    \item We develop a probing methodology that achieves near-complete coverage and construct a large-scale dataset comprising approximately \num{5.98} million active IPv6 addresses of Starlink user routers; this work was conducted under the guidance of Prof. Jianping Pan.
    \item We propose an efficient method to map Starlink’s backbone network using public IPv6 addresses, achieving coverage comparable to prior work while requiring only a limited number of probes.
    \item We conduct statistical analysis of Starlink users and PoPs in IPv6, investigate the relationship between Starlink PoPs and their served users, and characterize the IPv6 address assignment strategy of Starlink's gateways and PoPs. 
    \item We analyze the latency of Starlink user routers and uncover service unfairness among users within the same region. We further conduct follow-up measurements in July 2025 and observe clear performance improvements potentially driven by the ongoing V2 Mini satellite deployment.
\end{itemize}
}

To the best of our knowledge, this is currently the largest-scale measurement on the Starlink IPv6 network to date. To support and promote future research, we have made the dataset of the Starlink backbone network publicly available at \url{https://ki3.org.cn/\#/starlink-network}. In addition, the dataset of active Starlink IPv6 user routers is available at \url{https://ki3.org.cn/\#/datasetCenter} under specific conditions. We hope that our work will promote research on LEO satellite networks.




%% file: 02-BACKGROUND.tex
\section{BACKGROUND AND RELATED WORK}

\subsection{Background}

Starlink, a consumer-targeted LEO satellite Internet constellation operated by SpaceX, follows a bent-pipe architecture in most cases~\cite{hauri2020internet}. Most of its satellites orbit at an altitude of above \num{550} km in the 53\textdegree\ inclination plane. As of December 2025, Starlink provides coverage in over \num{155} countries and serves more than \num{9} million subscribers~\cite{k1}.


Figure~\ref{fig: Topology of the Starlink network} shows the components of the Starlink network. Each Starlink subscriber is equipped with a dish. When the subscriber tries to connect to the Internet, the traffic is forwarded to the dish by the user router. Then, the dish, which functions as a transparent bridge invisible at the network layer, forwards the traffic to satellites via phased-array antennas in Ku-band. The Starlink satellites can connect to multiple terminals simultaneously with multiple antennas subdivided into beams~\cite{mohan2024multifaceted}. When the satellites receive the traffic from the dish, they relay it to the ground station (GS) via a Ka-band link. The Starlink network adopts a one-hop bent-pipe routing approach given that the dish and GS are both within the visible range of the same satellite. Otherwise, Inter-Satellite Links (ISLs) need to be used. The GS relays the traffic to Starlink PoPs through terrestrial connections, prioritizing locations near data centers~\cite{li2014willow, li2014exr}. Subsequently, the traffic is routed across Starlink's backbone network, which comprises multiple backbone routers. Finally, the traffic exits the Starlink network and reaches its destination via Internet Exchange Points (IXPs) or Internet Service Providers (ISPs).

\begin{table*}[t]
\centering
\begin{threeparttable}
	\caption{Starlink PoPs and associated user routers (as of December 29, 2025)}
	\label{tab: PoP distribution}
	\centering
        \small
	\begin{tabularx}{0.9\textwidth}{>{\centering\arraybackslash}p{2cm}
    >{\centering\arraybackslash}p{1.5cm}
    >{\centering\arraybackslash}p{2cm}
    >{\centering\arraybackslash}X
    >{\centering\arraybackslash}X
    >{\centering\arraybackslash}X
    }
		\toprule
		Continent & \# of PoPs & PoP ID & PoP Location & \# of Routers served & \# of Regions served\\
		\midrule
		\multirow{18}{*}{North America} 
        & \multirow{18}{*}{18} & dllstxx1 &Dallas& 321065 & 2 \\
        && ashnvax2 &Ashburn& 255764 & 3 \\
        && mmmiflx1 &Miami& 225712 & 19 \\
        && chcoilx1 &Chicago& 224553 & 2 \\
        && atlagax1 &Atlanta& 219478 & 1 \\
        && sttlwax1 &Seattle& 214961 & 6 \\
        && qrtomex1 &Querétaro& 168114 & 2 \\
        && tmpeazx1 &Tempe& 157379 & 2 \\
        && snjecax1 &San Jos\'{e}& 152034 & 1 \\
        && clgycan1 &Calgary& 138502 & 4 \\
        && mplsmnx1 &Minneapolis& 130498 & 4 \\
        && mntlcan1 &Montréal& 115488 & 7 \\
        && knsymox1 &Kansas City& 114965 & 1 \\
        && sltyutx1 &Salt Lake City& 109068 & 6 \\
        && nwyynyx1 &New York City& 107933 & 6 \\
        && lsancax1 &Los Angeles& 100299 & 5 \\
        && gtmygtm1 &Guatemala City& 95173 & 7 \\
        && dnvrcox1 &Denver& 90191 & 3 \\ \hline
		\multirow{7}{*}{South America} 
        & \multirow{7}{*}{7} & splobra1 &S\~ao Paulo& 311365 & 3 \\
        && bnssarg1 &Buenos Aires& 226082 & 3 \\
        && sntochl1 &Santiago& 171841 & 3 \\
        && bgtacol1 &Bogot\'a& 160293 & 13 \\
        && brsabra1 &Bras\'{\i}lia& 137570 & 2 \\
        && frtabra1 &Fortaleza& 118737 & 4 \\
        && limaper1 &Lima& 91093 & 5 \\ \hline
		\multirow{6}{*}{Europe} 
        & \multirow{6}{*}{6} & frntdeu1 &Frankfurt& 158208 & 23 \\
        && lndngbr1 &London& 151275 & 5 \\
        && wrswpol1 &Warsaw& 137664 & 10 \\
        && mlnnita1 &Milan& 127176 & 9 \\
        && mdrdesp1 &Madrid& 121018 & 8 \\
        && sfiabgr1 &Sofia& 87834 & 29 \\ \hline
		\multirow{6}{*}{Oceania} 
        & \multirow{6}{*}{6} & sydyaus1 &Sydney& 120663 & 6 \\
        && mlbeaus1 &Melbourne& 93295 & 1 \\
        && acklnzl1 &Auckland& 84236 & 9 \\
        && brseaus1 &Brisbane& 82773 & 2 \\
        && prthaus1 &Perth& 45640 & 1 \\
        && chrhnzl1 &Christchurch& 8079 & 1 \\ \hline
        \multirow{9}{*}{Asia} 
        & \multirow{9}{*}{9} & mnlaphl1 &Manila& 113868 & 6 \\
        && jtnaidn2 &Jakarta& 58805 & 1 \\
        && srbaidn2 &Surabaya& 37273 & 1 \\
        && sngesgp1 &Singapore& 29823 & 8 \\
        && tkyojpn1 &Tokyo& 15794 & 5 \\
        && dhkabgd1 &Dhaka& 2206 & 2 \\
        && dohaqat1 &Doha& 2093 & 3 \\
        && msctomn1 &Muscat& 771 & 1 \\
        && mmbiind1 &Mumbai& 5 & 1 \\ \hline 
        \multirow{3}{*}{Africa} 
        & \multirow{3}{*}{3} & lgosnga1 &Lagos& 133123 & 11 \\
        && jhngzaf1 &Johannesburg& 127575 & 10 \\
        && nrbiken1 &Nairobi& 78085 & 11 \\
        
		\bottomrule
	\end{tabularx}

\end{threeparttable}
\end{table*}

\subsection{Related work}

The research community has conducted measurements on the Starlink network in recent years. Some work \cite{michel2022first,ma2023network,mohan2024multifaceted} investigates the user-perceived performance, such as latency, throughput, and packet loss rate. They compare user-perceived network performance of Starlink with that of terrestrial networks or traditional satellite networks. The work in~\cite{izhikevich2024democratizing} probes publicly exposed Internet services in IPv4 from the outside to measure the customer latency. It shows how Starlink's complex network architecture impacts customer latency. The work in~\cite{pan2023measuring, pan2024measuring} characterizes Starlink’s IPv4 access and backbone networks. However, it requires at least one probe associated with each Starlink PoP to verify the mapped Starlink backbone network in~\cite{pan2023measuring}. In addition, if they do not have available dishes or an RIPE Atlas~\cite{ripe_atlas} probe associated with a certain Starlink PoP, they may need to seek help online by through platforms like Reddit~\cite{reddit_starlink}. \cite{tanveer2023constellations} reports that the Starlink network employs a global scheduler to assign satellites to user terminals every 15 seconds, and presents an approach to identifying the satellite allocation mechanism. However, its reliance on physical Starlink hardware, such as the need to construct obstruction maps and perform \textit{inside-out} measurements, limits scalability and geographic coverage, particularly in regions where user terminals are difficult to obtain.


%% file: 03-METHODOLOGY.tex
\section{STARLINK NETWORK MEASUREMENT}

First, we propose a methodology for discovering the IPv6 addresses of active Starlink user routers and enumerating Starlink PoPs. Then, we illustrate our method for mapping the Starlink backbone network. Finally, we perform statistical analyses based on measurement results, including the distribution of Starlink users, PoPs, users served by PoPs, and the IPv6 address assignment strategy. 


\begin{figure*}[t] 
	\centering 
	\includegraphics[width=1\textwidth]{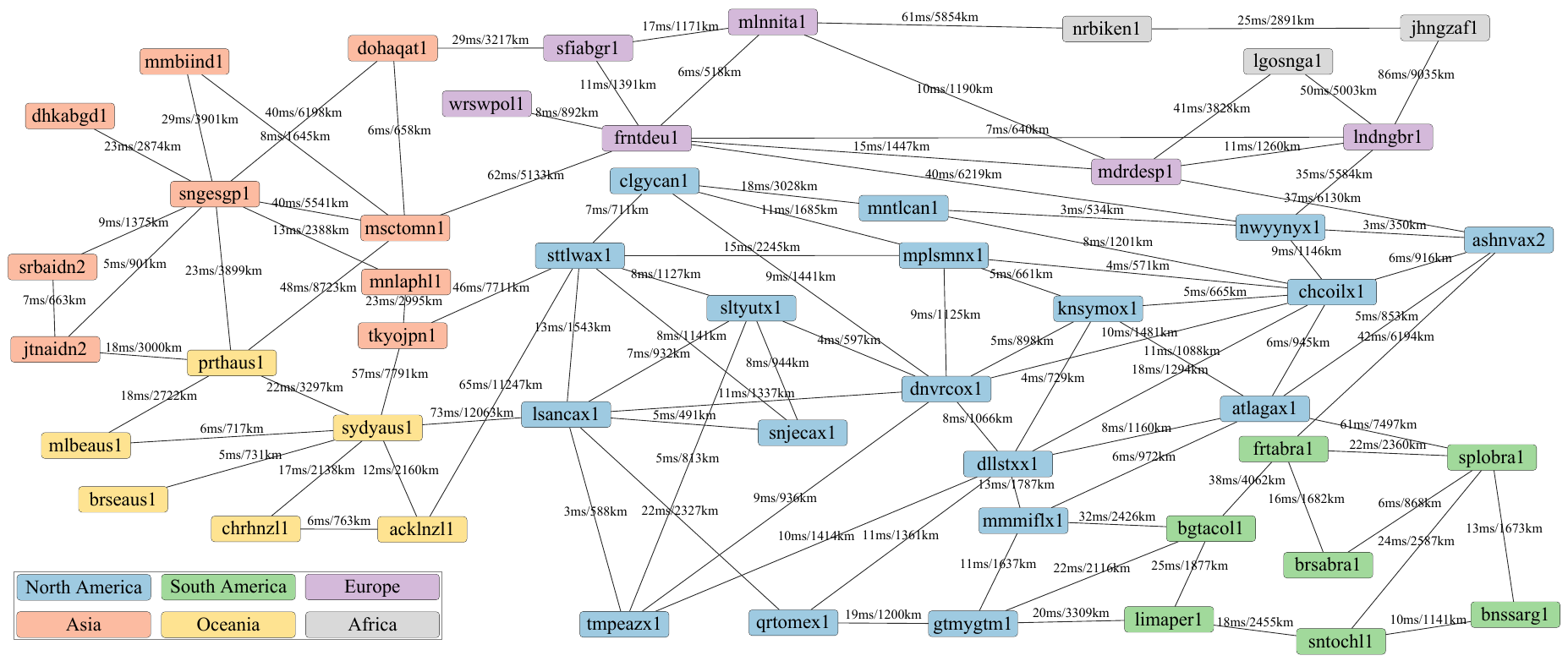} 
	\caption{Discovered Starlink backbone network (as of December 29, 2025).}
	\label{fig: Starlink backbone} 
\end{figure*}

\begin{figure}[t] 
	\centering 
	\includegraphics[width=0.45\textwidth]{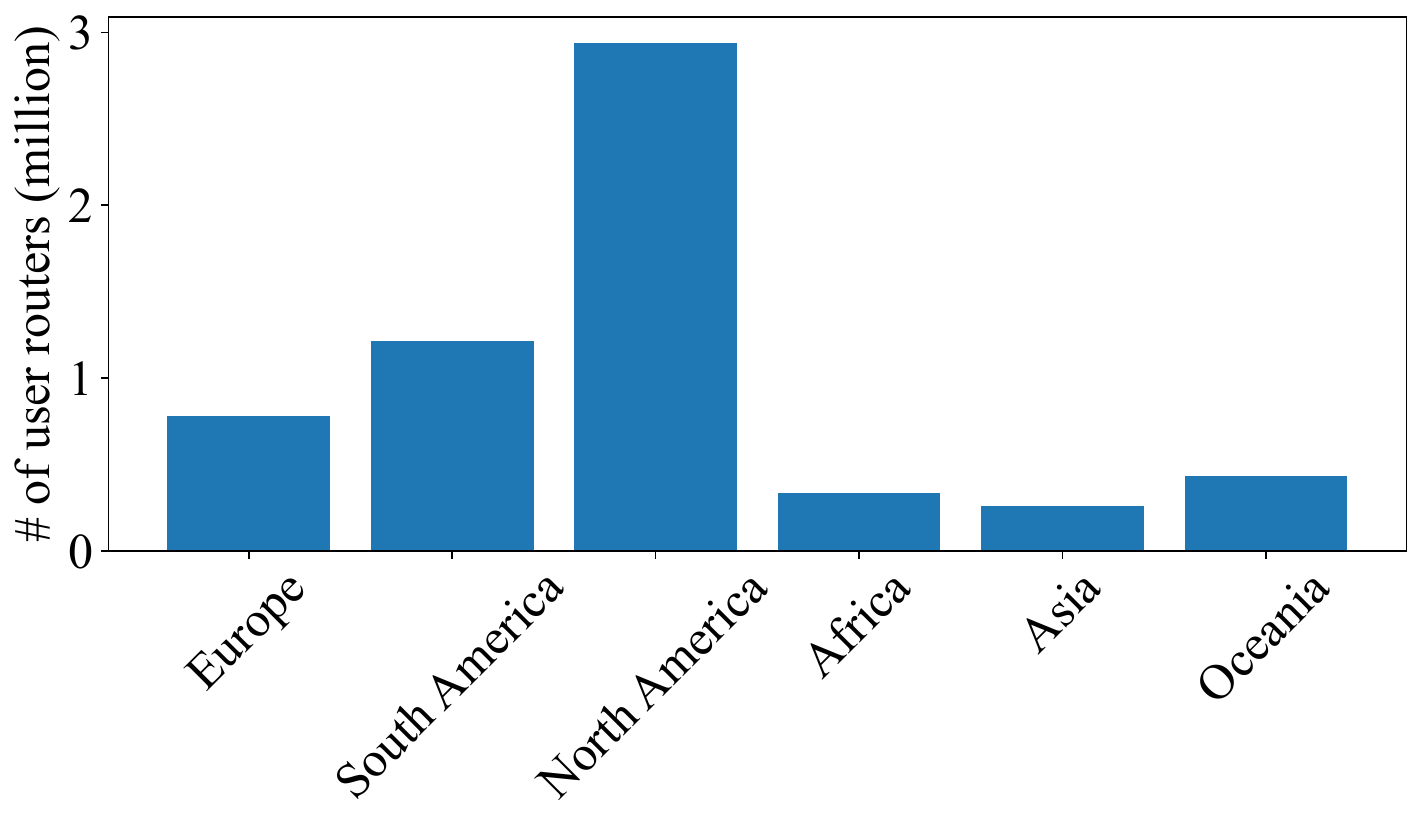} 
	\caption{Starlink User Routers per Continent (as of December 29, 2025).} 
	\label{fig: router per continent} 
\end{figure}

 \begin{figure}[t] 
 	\centering 
 	\includegraphics[width=0.45\textwidth]{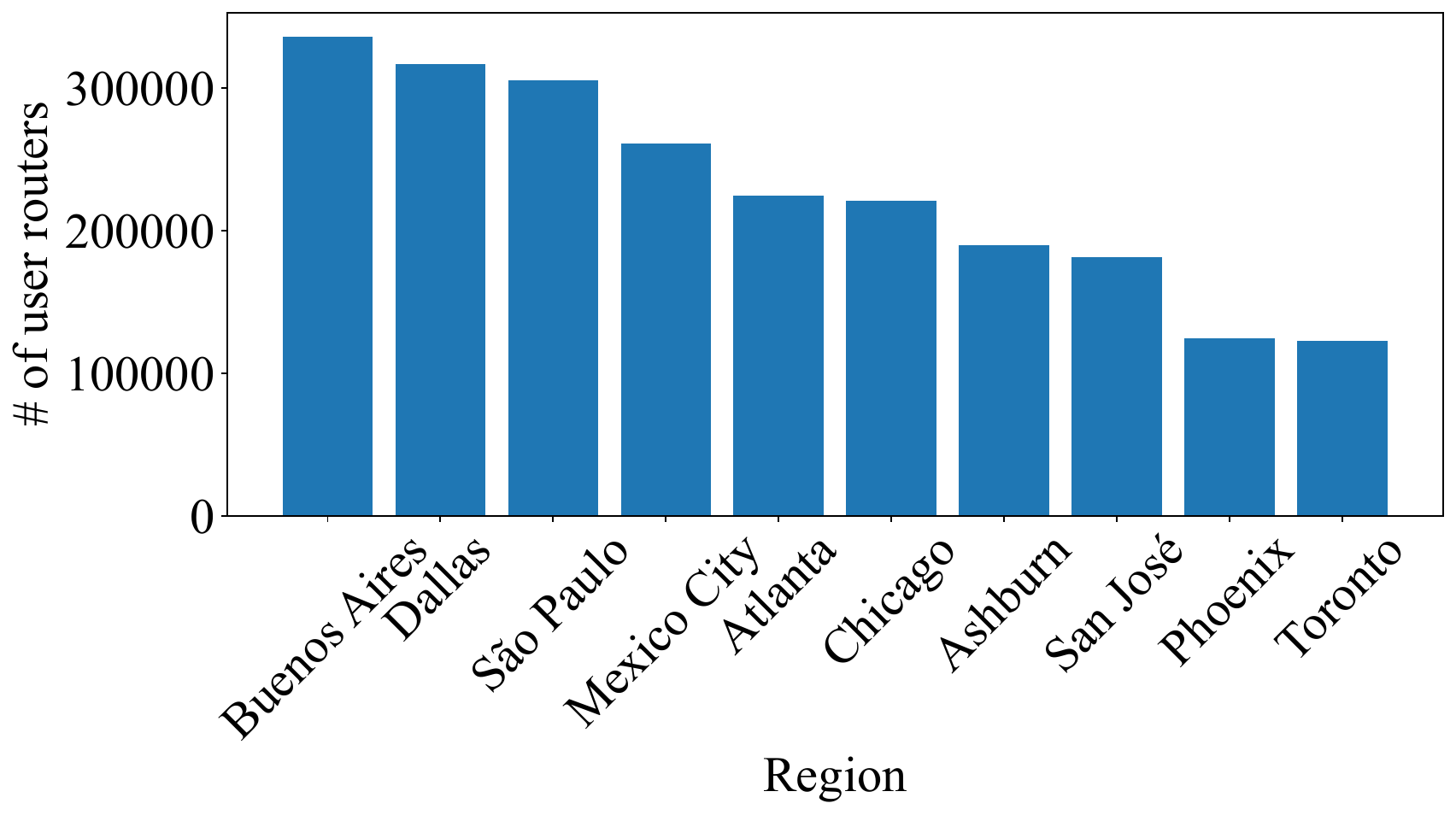} 
 	\caption{Top 10 Regions with the Most Starlink User Routers (as of December 29, 2025).} 
 	\label{fig: cities with most user routers} 
 \end{figure}

\subsection{Discovering Starlink user routers}

For regular Starlink subscribers, a CGNAT IPv4 address from the 100.64/10 subnet is assigned to the WAN side of the user router, with NAT performed at the associated home PoP. In the case of IPv6, each user router delegates a /56 prefix to its LAN clients~\cite{x1}. While GeoIP provides customer geolocations~\cite{p1}, we cannot probe every IP address with general-purpose IPv6 target generation algorithms due to the vast IP address space. 

Fortunately, we observe a regular pattern in the IPv6 addresses of Starlink user routers: The 57th to 127th bit values are set to 0, and the 128th bit is set to 1. For example, if Starlink users in a certain region are assigned /40 prefixes in the GeoIP database~\cite{p1}, we only need to focus on the remaining bits 41 to 56 to identify IPv6 addresses of active Starlink user routers. For this purpose, we generate candidates by enumerating different /56 sub-prefixes within each Starlink IPv6 prefix listed in the GeoIP feed, while fixing the remaining bits. For example, for Seattle, USA, the GeoIP database provides multiple prefixes, one of which is 2605:59c8::/40. We begin with 2605:59c8:00\textbf{00:00}\textcolor{gray}{00::1} (with the gray segments fixed) and then continue with \linebreak 2605:59c8:00\textbf{00:01}\textcolor{gray}{00::1}, up to 2605:59c8:00\textbf{ff:ff}\textcolor{gray}{00::1}.

This approach allows us to efficiently identify active Starlink user routers by scanning the generated candidates with XMap~\cite{xmap}, enabling the discovery of active IPv6 addresses associated with a specific geographic location. By using Starlink's address allocation rules. This approach exhaustively enumerates all possible targets, achieving more comprehensive and precise coverage than existing general-purpose IPv6 target generation algorithms, which may include redundant devices or miss active addresses. Finally, we identify \num{5975440} active Starlink user routers on December 29, 2025. Considering that some users may rely on IPv6-disabled routers or may have been offline during the measurement period, this number already accounts for more than 66\% of the total number of subscribers reported in~\cite{k1}, indicating a high level of coverage.


\subsection{Enumerating Starlink PoPs}

Starlink PoPs are globally distributed, and user traffic is tunneled to its home PoP before entering the public Internet. To identify the PoP associated with an active user router, we map its IPv6 address to the corresponding PoP based on Starlink’s official mapping of IP prefixes to PoPs~\cite{p2}.
Based on \num{5.98} million Starlink IPv6 addresses, we identify \num{49} Starlink PoPs as of December 29, 2025, as summarized in Table~\ref{tab: PoP distribution}. Compared to publicly available information from Starlink~\cite{peeringdb, googlemaps_starlink_pops}, we successfully identify all PoPs, achieving complete coverage.

\subsection{Mapping Starlink backbone network}

Starlink uses its backbone network to route user traffic between different PoPs. To investigate the Starlink backbone network, we need to determine the connections between these PoPs. For this purpose, we probe active user routers directly from a Starlink dish, as routing through other ISPs would bypass the backbone network.

First, we identify the IP addresses of the backbone routers within a Starlink PoP. When probing user routers from a Starlink dish, the traffic passes through the user router's home PoP and gateway before reaching the target user router. Therefore, we identify the third-to-last hop as the backbone router within the home PoP of the target user router. To determine the associated home PoP of the target user router, we map its IP address using Starlink’s official IP-to-PoP mapping.

However, a PoP may contain multiple backbone routers for performance reasons, meaning that data packets might pass through more than one hop within the same PoP. To identify additional backbone routers, we deduce that the routers belong to the same PoP if the latency between them is small (e.g.,$<3$ ms; results are insensitive to different thresholds). Starlink uses Multiprotocol Label Switching (MPLS) to reach its customers, as evidenced by the MPLS labels in traceroute results, so MPLS routers tunnel ICMP messages forward and then back, leading to inflated RTT readings for intermediate routers. To overcome this issue, we perform traceroutes directly to the backbone routers to obtain more stable latency measurements between backbone routers. In this way, we compile a complete list of backbone router IP addresses for each PoP.

Next, we deduce the connections between Starlink PoPs by the traceroute results from the dishes to the backbone routers. Since we have already established the relationship between backbone routers and PoPs, we can deduce the connections between PoPs by examining the connections between the backbone routers.

We utilize \num{2} probes from RIPE Atlas~\cite{ripe_atlas} and \num{4} Starlink dishes to perform traceroute measurements. The \num{2} RIPE Atlas probes are located in Southern Chile, and North Carolina, USA, respectively. 
The \num{4} dishes are located in Victoria, Canada; Brisbane, Australia; Bruehl, Germany; and Lusaka, Zambia.
As shown in Figure~\ref{fig: Starlink backbone}, we map the Starlink backbone network as of December 29, 2025, with the one-way delay and flight distance marked. It is important to note that, compared to Pan et al.~\cite{pan2023measuring}, we achieve comparable coverage using only a limited number of Starlink dishes as probes, without the need to place one in every PoP. The dataset of the Starlink backbone network is available at \url{https://ki3.org.cn/\#/starlink-network}.



\begin{table*}[t]
    \centering
    \caption{Probed regions and their associated data (July 2025)}
    \small
    \begin{tabularx}{\textwidth}{>{\centering\arraybackslash}p{4cm}|>{\centering\arraybackslash}p{5cm}|>{\centering\arraybackslash}p{3cm}|>{\centering\arraybackslash}X}
        Region & City displayed in GeoIP Feed & Home PoP ID& \# of active IP addresses\\ \hline
        American Samoa & Pago Pago & acklnzl1 & 290\\
        Samoa & Apia & acklnzl1 & 930\\
        Cook Islands & Avarua District & acklnzl1 & 1894\\
        Pitcairn Islands & Adamstown & acklnzl1 & 33\\
        Tuvalu & Funafuti & acklnzl1 & 682\\
        Nauru & Yaren & sydyaus1 & 821\\
        Tonga & Nuku'alofa & sydyaus1 & 1252\\
        U.S. Virgin Islands & Charlotte Amalie & mmmiflx1 & 3586\\
        Saint Martin (French part) & Marigot & mmmiflx1 & 828\\
        Saint Barthelemy & Gustavia & mmmiflx1 & 820\\
    \end{tabularx}
    \label{tab: Probed regions}
\end{table*}

\subsection{Statistics on measurement results}

\subsubsection{Distribution of Starlink user routers}
As shown in Figure~\ref{fig: router per continent}, we perform a statistical analysis of active Starlink IPv6 users distributed across different continents. The number of Starlink IPv6 users in North America reaches \num{2941177}, accounting for \num{49}\% of the total Starlink IPv6 users. Figure~\ref{fig: cities with most user routers} shows the top \num{10} regions with the most Starlink user routers. Although Buenos Aires has the largest number of users, the top 10 regions are dominated by regions in North America, with 8 out of 10 regions located there.


\subsubsection{Starlink PoPs and the users they serve}


Table~\ref{tab: PoP distribution} shows the Starlink PoPs and their served users. We observe that the Starlink Dallas PoP (dllstxx1) serves the most user routers, with a total of \num{321065}. Moreover, \num{8} PoPs serve more than \num{200000} user routers. Interestingly, we find that some PoPs serve users across multiple regions. For example, Starlink Sofia PoP (sfiabgr1) serves users in \num{29} regions in Europe, Asia, and Africa, and mmmiflx1 serves users in \num{19} regions in North and South America. In contrast, PoPs like Starlink Atlanta PoP (atlagax1) and Starlink Kansas City PoP (knsymox1) serve users only within their respective regions.

While Starlink users in most regions are served by a certain PoP, users in \num{53} regions are served by several PoPs. For instance, users in S\~ao Paulo are served by several PoPs, including Starlink Lima PoP (limaper1), Starlink S\~ao Paulo PoP (splobra1), Starlink Buenos Aires PoP (bnssarg1), Starlink Santiago PoP (sntochl1), and Starlink Bogotá PoP (bgtacol1). Users served by different PoPs are assigned different network prefixes in the same region. For example, users in S\~ao Paulo served by splobra1 are assigned 2803:9810:4300::/40, while those served by bgtacol1 are assigned 2803:9810:5380::/42.

\subsubsection{IPv6 address assignment strategy}

Based on the measurement results, we summarize the IPv6 address assignment strategy used in the Starlink network.

In the access network, the user router enables users to access the Internet. Each user router is assigned a public IPv6 address, with the 57th to 127th bits set to 0, and the 128th bit value set to 1. Devices connected to the user router are assigned IPv6 addresses within the same /56 prefix as the user router.

The IPv6 address of a Starlink gateway is structured as follows: the 49th to 64th bits range from 0x0248 to 0x0253, the 65th to 116th bits are set to 0, and the last 12 bits of the address range from 0x000 to 0x157. For example, a gateway may have an IPv6 address of 2620:134:b0fe:251::34. We also observe a correspondence between IPv4 and IPv6 addresses in the Starlink gateway system. Specifically, the IPv6 address of a gateway follows the pattern 2620:134:b0fe:x::y, where x and y correspond to segments of the IPv4 address 172.16.x.y. In particular, x and y are represented in hexadecimal IPv6 format but in decimal in the IPv4 format. For example, if a gateway's IPv6 address is 2620:134:b0fe:250::135, its corresponding IPv4 address is 172.16.250.135.

The IPv6 address blocks assigned to Starlink PoPs are 2620:134:b0ff::/116 and 2620:134:b004::/116.

%% file: 04-EXPERIMENTS.tex
\section{EXPERIMENT ON LATENCY in STARLINK}

In this section, we investigate the latency of Starlink user routers.\footnote{For this section, the data from the review period in July 2025 is preserved to prevent performance drifts caused by Starlink's rapid constellation evolution.} Based on a large dataset of active Starlink user routers' IPv6 addresses, we conduct TTL-limited ICMP probing to systematically analyze latency, latency variation, service unfairness, and longitudinal changes in Starlink’s performance.


\begin{figure*}[t] 
	\centering 
	\includegraphics[width=\textwidth]{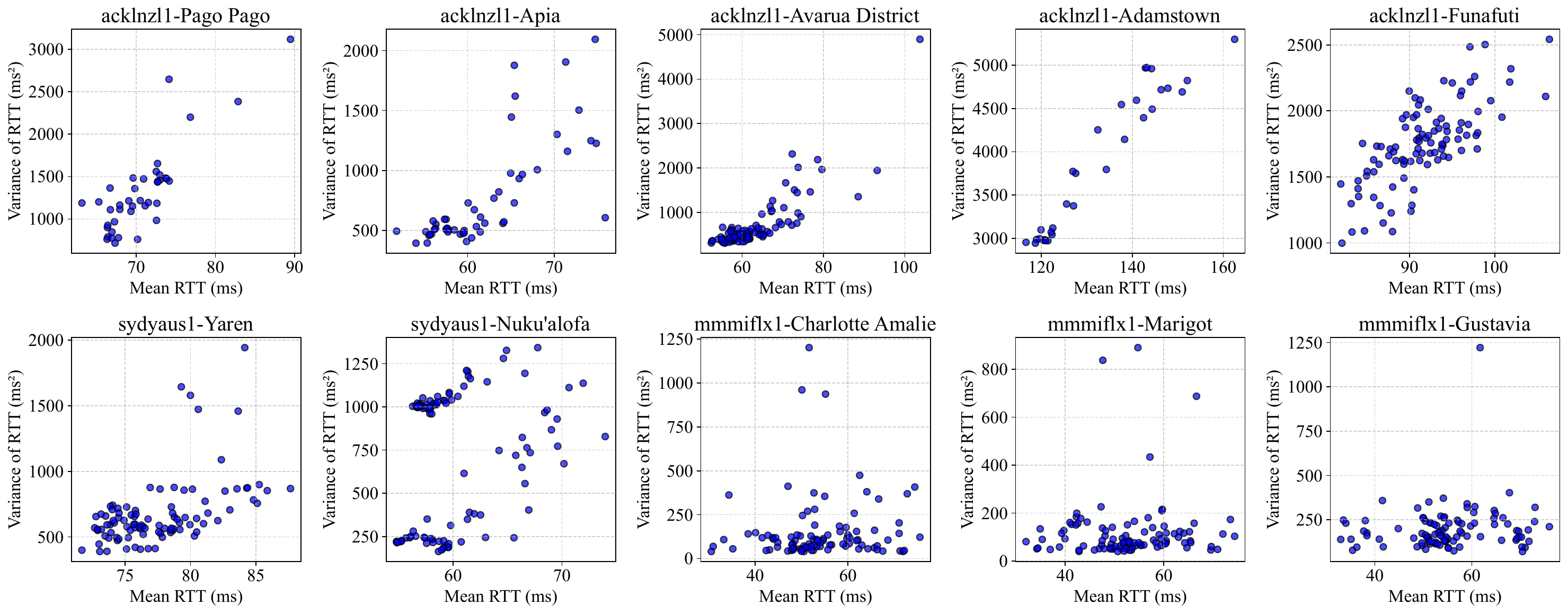} 
	\caption{Mean and variance of RTT for user routers by region (July 2025).} 
	\label{fig: mean-variance plot} 
\end{figure*}

\begin{figure}[t] 
	\centering 
	\includegraphics[width=0.49\textwidth]{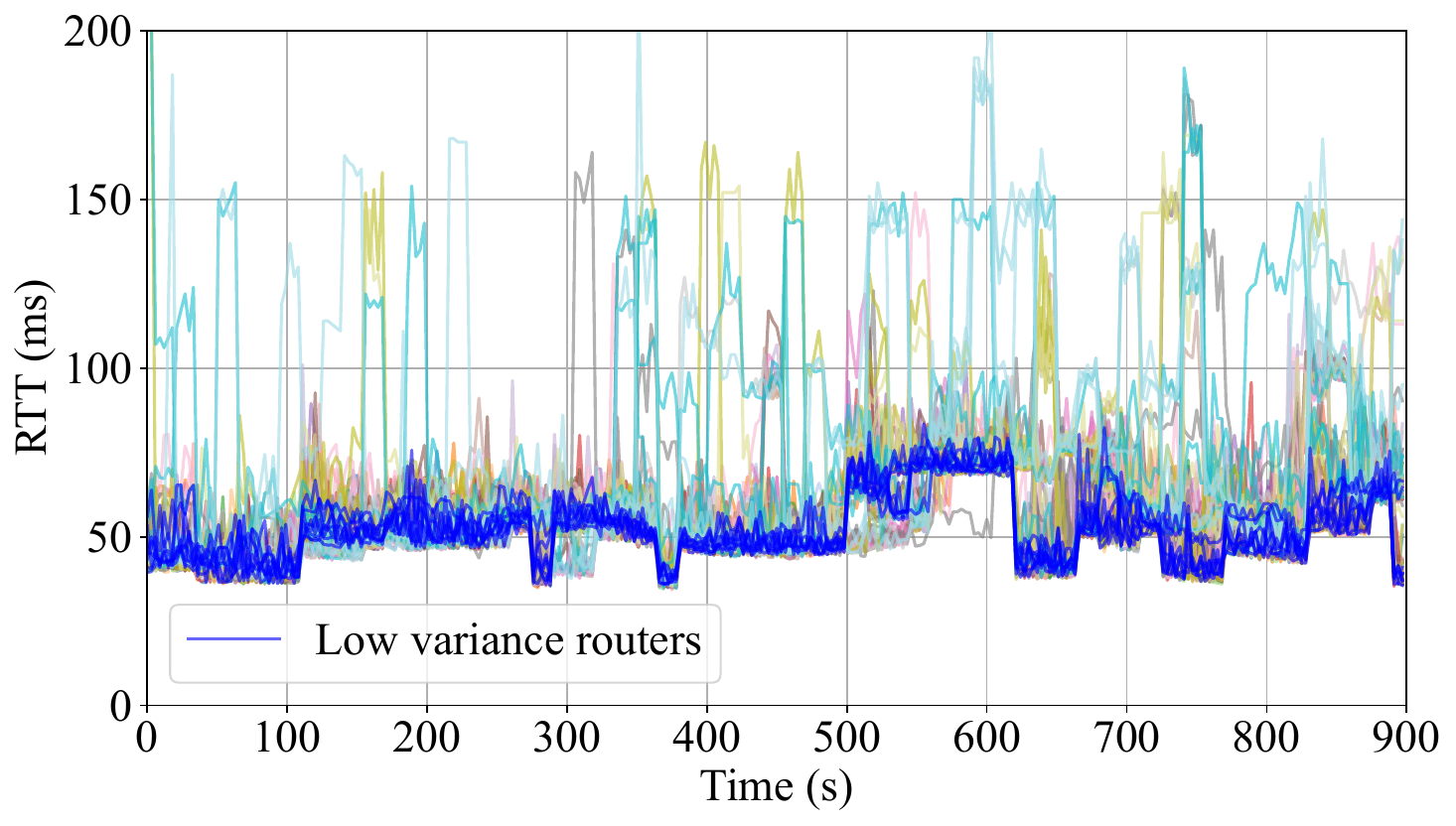} 
	\caption{RTT of user routers in Tonga (July 2025).} 
	\label{fig: Tonga latency group}
\end{figure}

\subsection{Experimental setup}

Since Starlink GeoIP Feed~\cite{p1} provides only region-level information, which typically represents large regions rather than a single city, we selected small island regions (e.g., Tonga and Nauru) as our probing targets to reduce GeoIP Feed uncertainty. Furthermore, the regions around our selected probing targets correspond to different entries in the Starlink GeoIP Feed, ensuring that the selected user routers are exclusively associated with these small island regions, thereby avoiding ambiguity from other regions. The details of these regions are summarized in Table~\ref{tab: Probed regions}.



We conducted measurements from a vantage point in Singapore on Alibaba Cloud to all selected target regions. To ensure that probe packets are forwarded via MPLS, we perform TTL-limited ICMP probing for each active IP address using traceroute. We send two ICMP echo requests with different TTL values to measure the latency to both the user router and its associated home PoP. Based on the topology shown in Figure~\ref{fig: Topology of the Starlink network}, where gateways and PoPs are typically co-located, the latency difference between these two points serves as an estimate of satellite hop latency, which we refer to simply as \emph{latency} in the remainder of this paper unless otherwise specified. For user routers whose home PoP is the Starlink Sydney PoP, the home PoP hop does not respond to our Singapore probes\footnote{This behavior was observed in our measurements conducted in May and July 2025. In more recent measurements, the Sydney PoP hop has become responsive to our probes.}. Accordingly, we utilize a virtual machine in Akamai's Sydney datacenter. Additionally, this server has good peering with the Starlink Sydney PoP, minimizing terrestrial network latency.

For each region, we randomly sample 100 IP addresses for probing (or use all available addresses if fewer than 100) and issue probes every 3 seconds. To avoid changes in the PoP hop identified by traceroute during prolonged measurements (e.g., due to routing dynamics or interface changes), we limit the probing window to one hour per region.

\subsection{Statistical characteristics of latency}

Figure~\ref{fig: mean-variance plot} shows the mean and variance of latency for user routers in different regions associated with Auckland, Sydney, and Miami PoPs. We use the mean latency rather than the minimum latency, since frequent jitter makes the latter insufficient to capture overall behavior. The results reveal substantial differences in latency variance among user routers associated with the Auckland and Sydney PoPs. In particular, for some regions, user routers with a higher latency variance also exhibit higher mean latency, resulting in a diagonal clustering pattern in the scatter plot.

Among the three PoPs analyzed, we focus primarily on the Sydney PoP for an in-depth investigation, while the Auckland- and Miami-associated regions are used only for statistical characterization. To investigate the source of the observed variance heterogeneity, we focus on user routers in Tonga as an illustrative example. We selected this region for two reasons: first, its user routers are associated with the Sydney PoP, where our vantage point maintains good peering to minimize terrestrial network interference, and second, Tonga has the highest number of active IPs among the small island regions served by Sydney PoP. Accordingly, we plot the time series of latency for these user routers in Figure~\ref{fig: Tonga latency group}. We group the routers into two categories based on the variance of their latency: Low-variance routers are colored blue, while high-variance routers are assigned distinct alternative colors. The rationale for the variance threshold used in this grouping will be explained in detail in Section \ref{subsec:latency-grouping}. 

The figure reveals that low-variance routers exhibit more stable connectivity, with fewer transitions between different latency levels. In contrast, high-variance routers frequently experience latency spikes, which persist for approximately 15 seconds (or multiples thereof) before returning to lower levels. This behavior is consistent with the handover interval of the global scheduler described in~\cite{tanveer2023constellations}. We hypothesize that low-variance routers are likely to correspond to users who are subscribed to the business plan.


In contrast, for the three regions served by the Miami PoP, the differences in latency variance among user routers are markedly smaller. This is likely due to the relatively dense deployment of ground stations in those regions. As a result, even if lower-priority users do not share the same network path as higher-priority users, the impact on their network latency remains minimal. Although there is still a difference between the average latency of low-priority and high-priority users, taking different network paths allows these routers to experience comparable connection stability.

\begin{figure}[t] 
	\centering 
	\includegraphics[width=0.5\textwidth]{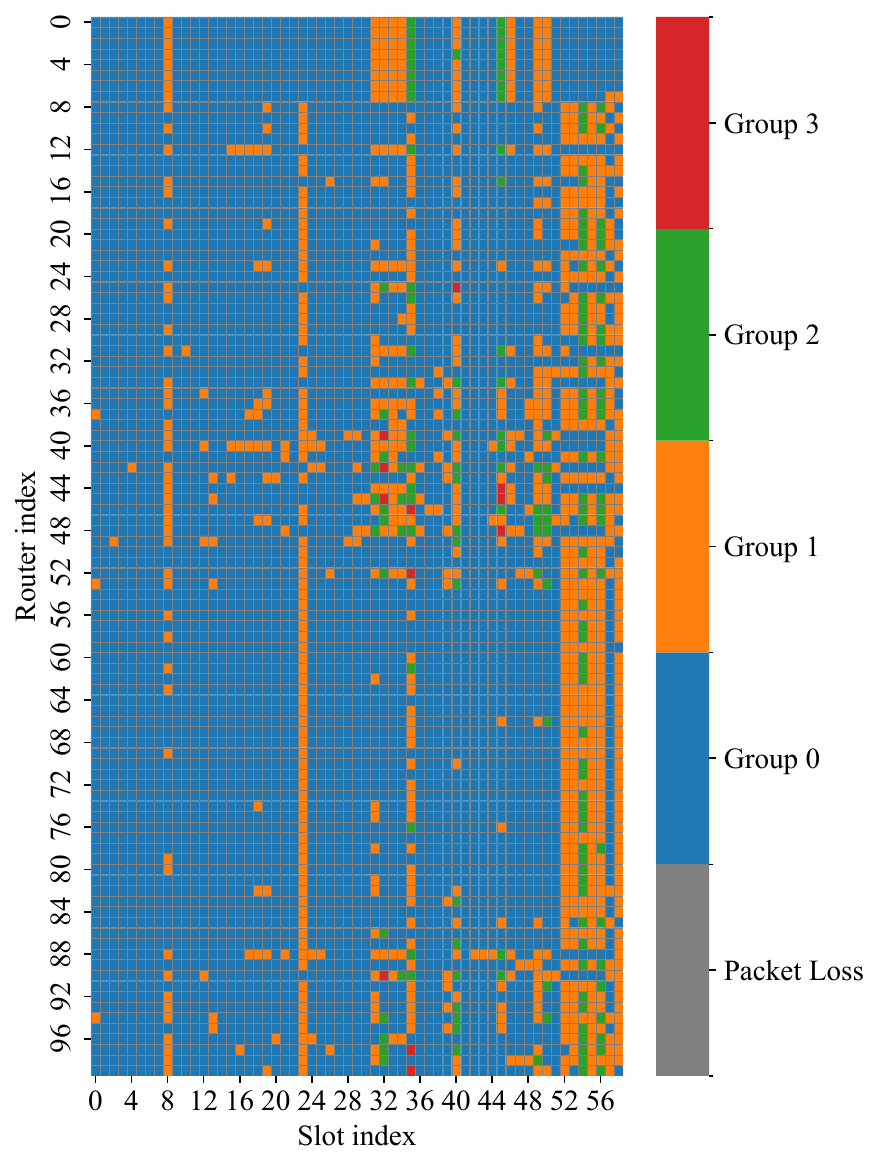} 
	\caption{Evolution of latency groups for user routers in Tonga (May 2025).}
	\label{fig: group plot may}
\end{figure}

\subsection{Latency grouping and service unfairness}
\label{subsec:latency-grouping}
In the previous subsection, we empirically confirmed that Starlink operates under a globally synchronized 15-second scheduling cycle, as described in~\cite{tanveer2023constellations}. Although Starlink now supports dynamic beam switching, we assume that user dishes are free of obstructions. Under this assumption, satellite handovers are less likely to occur within a 15-second scheduling window than at scheduling boundaries. This allows us to attribute differences in latency primarily to variations in routing paths from the user router to the PoP. Therefore, we divide the timeline into discrete 15-second slots, each representing a single scheduling cycle.

To further analyze the latency within each slot, we introduce the concept of latency groups, each of which corresponds to a group of routers with similar minimum latency. 
Figure~\ref{fig: Tonga latency group} shows that the latency within each slot can be clustered into multiple groups, which likely reflect different routing paths from the dish to the PoP, as discussed in~\cite{izhikevich2024democratizing}.
We infer these group assignments by applying $k$-means clustering to the minimum latency of each router within a given slot. We treat each slot independently, as our primary focus is the correlation of group assignments among routers.
The optimal number of clusters is chosen using the Silhouette score~\cite{rousseeuw1987silhouettes}, with an added regularization term to mitigate overfitting.

This yields the estimated latency group of each router in each time slot, as illustrated in Figure~\ref{fig: group plot may}, where user routers in Tonga are displayed and ordered by their overall latency variance for visual clarity. We observe that the eight routers with the lowest variance remain the same latency group for the vast majority of the measurement period, with only occasional deviations. In contrast, this consistency is not observed among other routers. This suggests that these routers may receive higher priority in resource allocation, which is consistent with our earlier hypothesis that they correspond to users subscribed to the business plan.

Based on this observation, we find that the correlation of latency group assignments over time could offer a practical basis for empirically determining a threshold for latency variance. This potentially enables a fully \textit{outside-in} method for distinguishing between user types within the same geographic region.

\begin{figure}[t] 
	\centering 
	\includegraphics[width=0.5\textwidth]{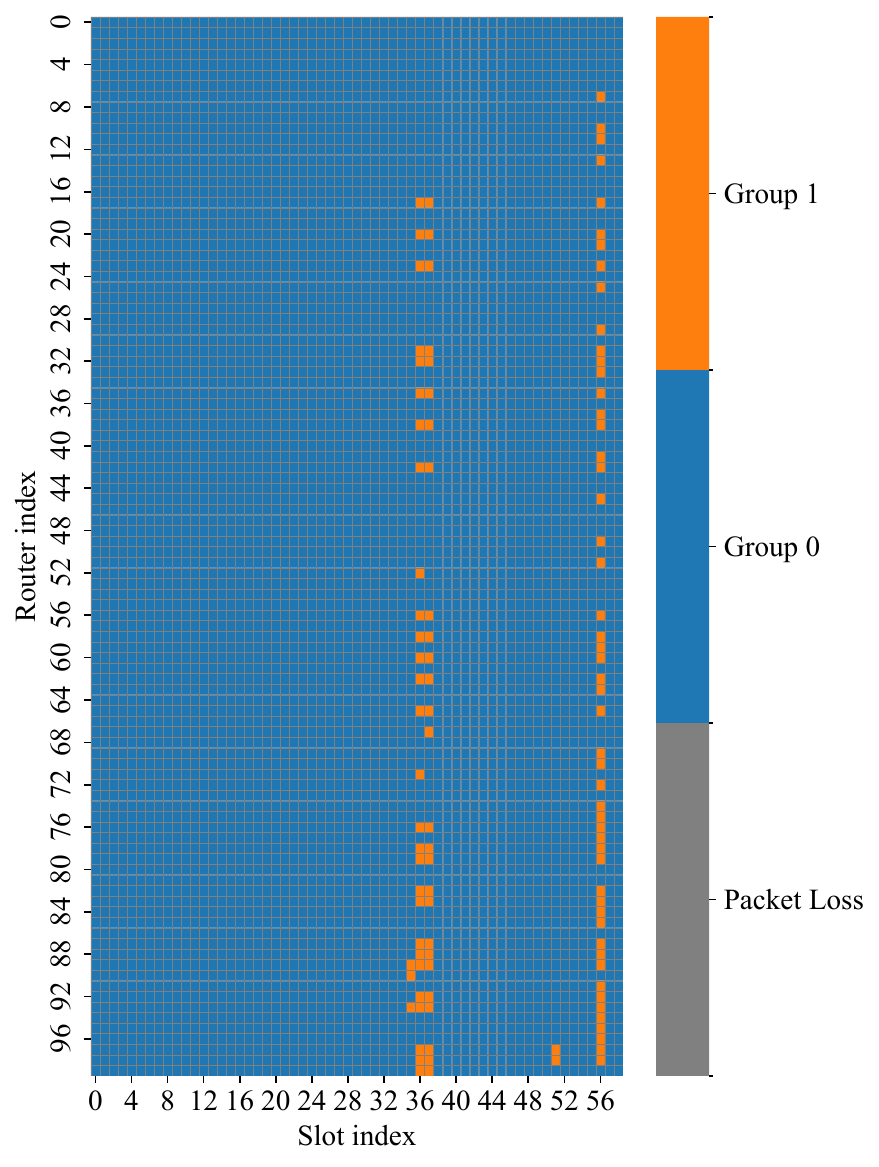} 
	\caption{ Evolution of latency groups for user routers in Tonga (July 2025).}
	\label{fig: group plot july}
\end{figure}

\begin{table}[t]
    \centering
    \caption{Mean latency (ms) by region (May vs. July 2025)}
    \small
    \begin{minipage}{0.46\textwidth}
    \begin{tabularx}{\linewidth}{>{\centering\arraybackslash}X|>{\centering\arraybackslash}p{1.75cm}|>{\centering\arraybackslash}p{1.75cm}}
        Region & Mean latency (May) & Mean latency (July)\\ \hline
        American Samoa & 71 & 54\\
        Samoa & 62 & 50\\
        Cook Islands & 63 & 56\\
        Pitcairn Islands & 133 & 87\\
        Tuvalu & 92 & 59\\
        Nauru & 78 & 62\\
        Tonga & 60 & 55\\
        U.S. Virgin Islands & 54 & 43\\
        Saint Martin (French part) & 53 & 49\\
        Saint Barthelemy & 55 & 47\\
    \end{tabularx}
    \end{minipage}
    \label{tab:mean latency}
\end{table}

\subsection{Longitudinal changes in Starlink's latency}

To investigate the longitudinal evolution of Starlink's performance, we conducted measurements across all regions in July 2025. As shown in Table~\ref{tab:mean latency}, all probed regions experienced a substantial decrease in average latency.

To further explore the underlying cause of this improvement, we re-visualize the estimated latency group of each router in each time slot, as described in Section~\ref{subsec:latency-grouping}. Evolution of latency groups for user routers in Tonga are shown in Figure~\ref{fig: group plot july}. The results reveal a significant reduction in routing path divergence (i.e., co-located user routers traversing different paths to the PoP at the same time slot). This observation suggests that the latency improvement cannot be solely attributed to general network optimizations such as software updates. It also reflects an improvement in the load capacity of the network link, as fewer flows are routed to take alternative paths.

A possible explanation for this improvement is the large-scale deployment of new-generation satellites. Specifically, 731 V2 Mini satellites have been launched during this period, which accounts for about 17\% of the total number of V2 Mini satellites in orbit~\cite{wikipedia_starlink_launches,planet4589_starlink_stats}. Each V2 Mini satellite offers a bandwidth of 96 Gbps, quadrupling the 24 Gbps capacity of the previous V1.5 satellites~\cite{starlink_progress_2024}. The deployment of V2 Mini satellites thus acts as a potential driver of the observed decline in path variability and the overall reduction in latency.

In summary, our dataset provides sufficiently broad coverage of Starlink routers across regions, thereby enabling a systematic study of the evolution of Starlink’s network latency performance.

%% file: 05-DISCUSSION.tex
\section{DISCUSSION}

\subsection{Limitations}

\noindent\textbf{User geolocation inaccuracy issues.} We use the Starlink GeoIP database~\cite{p1} to geolocate user routers. However, a single location entry may correspond to a geographic area much larger than a single city. For example, the location of Lagos may represent a wide region in Africa. We therefore focus on small island regions in this study. Finer-grained geolocation data could enable characterization of a broader Starlink user base.



\noindent\textbf{Uneven distribution of available Starlink dishes.} Most of our measurement nodes are located in Europe and the U.S., which may cause some inter-regional connections to be missed. For example, while U.S.–Asia and U.S.–Europe connections can be observed, Asia–Europe connections may not be captured. Expanding the global distribution of Starlink dishes would enable more comprehensive mapping of the backbone network.

\noindent\textbf{Limitations of outside-in measurement accuracy.} \textit{Outside-in} measurements are less precise than \textit{inside-out} measurements because they lack synchronized time sources (e.g., NTP) and access to internal data (e.g., obstruction maps). Consequently, the exact one-way delay and routing paths cannot be determined definitively. 


\subsection{Ethics}

In this paper, we identify approximately \num{5.98} million active Starlink IPv6 user routers and map the backbone network based on the measurement results. We strictly limit the aggregate probing rate to 8.32 Mbps and send at most one packet every 3 seconds to each user router, ensuring negligible impact on network performance. In addition, all probes originate from fixed source IP addresses, allowing users to easily identify and block our traffic. We protect user privacy throughout the measurement process. To minimize exposure and ensure data security, only the IP addresses of backbone routers are made publicly available. Researchers or organizations interested in the active IPv6 addresses of Starlink user routers may access the dataset upon request and with a clearly stated purpose. No other aspects of this work raise ethical concerns.

%% file: 06-CONCLUSION.tex
\section{CONCLUSION}
In this paper, we conduct the largest-scale measurement of the Starlink IPv6 network to date. We obtain a comprehensive list of IPv6 addresses of active user routers via a probing methodology with near-complete coverage. Furthermore, we map the Starlink backbone network based on the measurement results. Specifically, we show that our proposed methodology for mapping the backbone network is highly efficient. In addition, we perform a statistical analysis of Starlink routers, PoPs, and the IPv6 assignment strategy. We investigate the latency of the Starlink user routers and identify service unfairness among users in the same region. We also observe improvements in latency over time, which may be partially attributed to the ongoing deployment of V2 Mini satellites. Our work fills a gap in the Starlink IPv6 network measurement, and we hope that it will help deepen the community's understanding of LEO satellite networks, promoting research and accelerating their development.
